\documentclass[
aps,
nofootinbib,
prd,
superscriptaddress,
tightenlines,
notitlepage,
twocolumn,
showpacs,
floatfix
]{revtex4-1}

\usepackage[colorlinks]{hyperref}
\usepackage{tabularx}
\usepackage{graphicx}
\usepackage{amssymb,amsmath,bm,tensor,braket,mathtools}
\usepackage{xcolor}
\usepackage[varg]{txfonts}
\usepackage{enumerate}
\usepackage{mathtools}
\usepackage[capitalize]{cleveref}
\usepackage[utf8]{inputenc}
\usepackage[normalem]{ulem}

\def\al{\alpha}
\def\be{\beta}

\def\de{\delta}
\def\ep{\epsilon}

\def\th{\theta}

\def\la{\lambda}

\def\rh{\rho}

\def\si{\sigma}

\def\ph{\phi}

\def\om{\omega}
\def\Ga{\Gamma}
\def\De{\Delta}

\def\Ph{\Phi}

\def\Om{\Omega}

\def\mn{{\mu\nu}}

\def\lsim{\mathrel{\rlap{\lower4pt\hbox{\hskip1pt$\sim$}}
    \raise1pt\hbox{$<$}}}
\def\gsim{\mathrel{\rlap{\lower4pt\hbox{\hskip1pt$\sim$}}
    \raise1pt\hbox{$>$}}}
\def\sqr#1#2{{\vcenter{\vbox{\hrule height.#2pt
         \hbox{\vrule width.#2pt height#1pt \kern#1pt
         \vrule width.#2pt}
         \hrule height.#2pt}}}}

\def\prt{\partial}
\def\lrpartial{\raise 1pt\hbox{$\stackrel\leftrightarrow\partial$}}

\def\part2{\partial_\alpha \partial^\alpha}

\def\xx'{|\vec x -\vec x'|}

\def\b2{b^\al b_\al}

\newcommand{\beq}{\begin{equation}}
\newcommand{\eeq}{\end{equation}}
\newcommand{\bea}{\begin{eqnarray}}
\newcommand{\eea}{\end{eqnarray}}
\newcommand{\bit}{\begin{itemize}}
\newcommand{\eit}{\end{itemize}}
\newcommand{\rf}[1]{(\ref{#1})}
\newcommand\bw{\begin{widetext}}
\newcommand\ew{\end{widetext}}

\newcommand{\IMM}{\affiliation{Institute for Multidisciplinary Mathematics, Polytechnic University of Val\`{e}ncia, Val\`{e}ncia 46022, Spain}}

\newcommand{\BIMSA}{\affiliation{Beijing Institute of Mathematical Sciences and Applications, Beijing 101408, China}}

\newcommand{\THU}{\affiliation{Department of Astronomy,
Tsinghua University, Beijing 100084, China}}

\begin{document}

\title{The I-Love universal relation for polytropic stars under Newtonian gravity}

\date{\today}
\author{Rui Xu}\email[Corresponding author:]{xuru@tsinghua.edu.cn} \THU
%\author{Dandan Xu}\THU
\author{Alejandro Torres-Orjuela}\BIMSA
\author{Lars Andersson}\BIMSA
\author{Pau Amaro Seoane}\IMM

\begin{abstract} 
%The moment of inertia and the tidal deformability of idealized stars with polytropic equations of state (EOSs) are numerically calculated in both Newtonian gravity and general relativity (GR). The results explicitly verify that the relation between the moment of inertia and the tidal deformability, when parameterized by the mass of the star, varies within $1\,\%$ and $10\,\%$ for different polytropic indices in Newtonian gravity and in GR separately, indicating a better I-Love universal relation in the Newtonian case. The theoretically calculated I-Love universal relation for the polytropic stars is then tested against measured data of the moment of inertia and the tidal deformability for the 8 planets and some of their moons in our solar system. We find that the theoretical I-Love universal relation matches the observational data well, so it can be used as an empirical relation for estimating the moment of inertia or the tidal deformability of an exoplanet if one of the two quantities and the mass of the exoplanet is known.    

The moment of inertia and tidal deformability of idealized stars with
polytropic equations of state (EOSs) are numerically calculated under both
Newtonian gravity and general relativity (GR). The results explicitly confirm
that the relation between the moment of inertia and tidal deformability,
parameterized by the star's mass, exhibits variations of $1\,\%$ to $10\,\%$
for different polytropic indices in Newtonian gravity and GR, respectively.
This indicates a more robust I-Love universal relation in the Newtonian
framework. The theoretically derived I-Love universal relation for polytropic
stars is subsequently tested against observational data for the moment of
inertia and tidal deformability of the 8 planets and some moons in our solar
system. The analysis reveals that the theoretical I-Love universal relation
aligns well with the observational data, suggesting that it can serve as an
empirical relation. Consequently, it enables the estimation of either the
moment of inertia or the tidal deformability of an exoplanet if one of these
quantities, along with the mass of the exoplanet, is known.

\end{abstract}

\maketitle
\allowdisplaybreaks  % allow long equations to display on different pages
 
%---------------------------------------------------------------------
\section{Introduction}
\label{sec:intro}

The universal relations for neutron stars (NSs) and quark stars (QSs) were
first identified by Yagi and Yunes in terms of the moment of inertia (I), tidal
deformability (Love), and quadrupole moment (Q) of these compact stars
\cite{Yagi:2013bca, Yagi:2013awa}. These relations, known as I-Love-Q universal
relations, are termed ``universal'' because they exhibit remarkable insensitivity
to the equation of state (EOS) of the compact object. Even for EOSs that
produce significantly different mass-radius relations for NSs and QSs, the
I-Love-Q relations hold with a precision better than $1\%$. Subsequently, the
universal relations were extended to include additional perturbative properties
of compact stars, such as higher-order multipolar tidal deformabilities
\cite{Yagi:2013sva,Yagi:2014bxa,Pani:2015nua,Delsate:2015wia,Yagi:2016bkt} and
oscillation mode frequencies
\cite{Chan:2014kua,Chirenti:2015dda,Doneva:2015jba}. While these extended
universal relations are valuable, their precision is generally lower than that
of the original I-Love-Q relations, with discrepancies reaching up to $10\%$
for different EOSs \cite{Maselli:2013mva}.

The precise mechanism underlying the universality of these relations remains
unclear. Unlike the no-hair theorem for black holes in general relativity (GR)
\cite{Bekenstein:1995un}, the I-Love-Q universal relations are not exact and
are established through numerical studies. Their accuracy often deteriorates
when simplifying assumptions used in structural and perturbative calculations
of stars are relaxed \cite{Taylor:2019hle,Chen:2023bxx}. For instance, factors
such as anisotropic pressure, non-barotropic EOSs, or the inclusion of more
complex physics introduce greater discrepancies in the universal relations
\cite{Silva:2014fca,Yagi:2015hda,Yagi:2016bkt,Martinon:2014uua,Yeung:2021wvt,Laskos-Patkos:2023vre,Guedes:2024gxo}.
Moreover, the validity of the universal relations can also depend on the
gravitational theory. While the I-Love universal relation has been shown to
hold in many scalar-tensor theories (albeit with up to $10\%$ discrepancies)
\cite{Doneva:2017jop,Gupta:2017vsl,Hu:2021tyw}, it is visibly broken in
specific scalar-tensor theories where the Gauss-Bonnet invariant is coupled to
the square of a scalar field \cite{Xu:2021kfh}. In such cases, the
Schwarzschild solution is no longer unique, and scalarized black holes appear,
violating the no-hair theorem
\cite{Silva:2017uqg,Doneva:2017bvd,Antoniou:2017acq}. Whether a connection
exists between the breakdown of the I-Love universal relation and the violation
of the no-hair theorem warrants further investigation.

Although the universal relations for NSs and QSs are widely studied because
they mitigate EOS-related ambiguities in compact star studies, they are not
exclusive to compact stars. As shown analytically in Ref.~\cite{Yagi:2013awa}
using two specific polytropic EOSs, the I-Love-Q universal relations also apply
to spherical objects in the Newtonian limit. This work explicitly confirms the
I-Love universal relation for general polytropic EOSs in Newtonian gravity and
compares it with the corresponding relation in GR. A noteworthy application of
the I-Love universal relation for polytropic EOSs in Newtonian gravity is its
use in analyzing the tidal Love number and moment of inertia of solar-system
planets and their moons. Despite their layered structures and complex
compositions, which deviate from simple polytropic EOSs, these celestial
objects conform well to the I-Love universal relation. This finding highlights
the potential of the I-Love relation as a powerful tool not only for studying
compact stars but also for deducing properties of exoplanets.

%If two perturbation properties of NSs involved in a universal relation are measured, for example, moment of inertia and tidal deformability of a NS simultaneously measured in a multi-messenger binary coalescence observation in the future, then the measurements can be used to distinguish gravity theories that give different universal relations between the two perturbation properties.   

%We start with reviewing the formulae to calculate the tidal deformability and the moment of inertia in Newtonian gravity in Sec.~\ref{sec:IIa} and in GR in Sec.~\ref{sec:IIb}. In Sec.~\ref{sec:IIc}, we show the numerical results for polytropic EOSs. In Sec.~\ref{sec:checking}, measurements of the tidal Love number and the moment of inertia for 17 planets and moons are listed and plotted against the I-Love universal relation for polytropic stars. We give a brief summary in Sec.~\ref{sec:sum}. It is interesting to see how the I-Love relation for white dwarfs deviates a little from the I-Love universal relations in both Newtonian gravity and GR. So we put relevant results in Appendix~\ref{sec:ILwd}. Geometrized units where the gravitational constant $G$ and the speed of light $c$ are equal to one have been used throughout the work. SI units are also used to express values of physical quantities. The sign convention of the metric is $(-, +, +, +)$. 
%

We begin by reviewing the formulae used to calculate the tidal deformability
and the moment of inertia in Newtonian gravity in Sec.~\ref{sec:IIa}, and in
general relativity (GR) in Sec.~\ref{sec:IIb}. In Sec.~\ref{sec:IIc}, we
present numerical results for polytropic EOSs. In Sec.~\ref{sec:checking}, we
list and plot measurements of the tidal Love number and the moment of inertia
for 17 planets and moons, comparing them with the I-Love universal relation for
polytropic stars. A brief summary is provided in Sec.~\ref{sec:sum}. 

It is worth noting that the I-Love relation for white dwarfs deviates slightly
from the I-Love universal relations in both Newtonian gravity and GR. The
relevant results are included in Appendix~\ref{sec:ILwd}. Throughout this work,
geometrized units are used, where the gravitational constant $G$ and the speed
of light $c$ are set to one. SI units are also employed to express the values
of physical quantities. The metric sign convention used is $(-, +, +, +)$.

%---------------------------------------------------------------------
\section{I-Love universal relations for polytropic stars}
\label{sec:II}
%---------------------------------------------------------------------

\subsection{Newtonian gravity}
\label{sec:IIa}

A comprehensive account of the spherically hydrostatic stars with polytropic EOSs and their perturbations can be found in Ref.~\cite{Poisson:2014}. We briefly review the formulae here. First, the basic set of equations consists of the Poisson's equation and the energy-momentum conservation equations,
\bea
&& \nabla^2 \Ph = -4\pi \rh ,
\nonumber \\
&& \frac{\prt \rh}{\prt t} + \nabla \cdot \left(\rh {\boldsymbol{v}} \right) = 0, 
\nonumber \\
&& \frac{\prt {\boldsymbol{v}}}{\prt t} + \left( {\boldsymbol{v}} \cdot \nabla \right) {\boldsymbol{v}} = \nabla \Ph - \frac{1}{\rh}\nabla p ,
\label{basiceqs1}
\eea   
where $\Ph$ is the Newtonian potential, and $\rh, \, p, \, {\boldsymbol{v}}$ are the density, the pressure, and the velocity of the fluid constituting the star. Then, the variables assume the form 
%\ATO{Do you mean `We assume for the variable the form'? I do not think the variables have a consciousness to assume anything:D} \textbf{Pau: In English, as it is the case of Spanish (asumir, adquirir), "to assume" also means "to adopt", so that the sentence is ok.}
\bea
&& \Ph = \Ph^{(0)} + \de \Ph, 
\nonumber \\
&& \rh = \rh^{(0)} + \de \rh, 
\nonumber \\
&& p = p^{(0)} + \de p,
\nonumber \\
&& {\boldsymbol{v}} = {\boldsymbol{v}}^{(0)} + \de {\boldsymbol{v}} .
\label{varexp1}
\eea
%with the zeroth-order terms corresponding to the spherically hydrostatic solution and the perturbation terms referring to the tidal perturbation in our consideration. The zeroth-order terms of the variables only depend on the radial coordinate $r$ with ${\boldsymbol{v}}^{(0)}=0$ especially, so one obtains the hydrostatic equations for spherical Newtonian stars,\ATO{Maybe: `If one further assumes the zeroth-order terms of the variables to only depend on the radial coordinate $r$ with ${\boldsymbol{v}}^{(0)}=0$ especially, one obtains the hydrostatic equations for spherical Newtonian stars'}
The zeroth-order terms correspond to the spherically symmetric hydrostatic solution, while the perturbation terms account for the tidal perturbation under consideration. If one assumes that the zeroth-order terms of the variables depend only on the radial coordinate $r$, with ${\boldsymbol{v}}^{(0)}=0$, the hydrostatic equations for spherical Newtonian stars are obtained:
%\ATO{Alternatively: "If one further assumes the zeroth-order terms of the variables to only depend on the radial coordinate $r$ with ${\boldsymbol{v}}^{(0)}=0$ especially, one obtains the hydrostatic equations for spherical Newtonian stars."}
\bea
&& \Ph^{(0)\, \prime\prime} + \frac{2}{r} \Ph^{(0)\, \prime} = -4\pi \rh, 
\nonumber \\
&& p^{(0)\, \prime} = \rh^{(0)} \Ph^{(0)\, \prime} ,
\label{bgeq1}
\eea 
where the prime denotes the derivative with respect to $r$. 

For the perturbation of Eq.~\rf{basiceqs1}, it is convenient to use the Lagrangian displacement ${\boldsymbol{\xi}}$ of the perturbed fluid elements. Without considering the uninteresting odd-parity perturbations, ${\boldsymbol{\xi}}$ can be expressed in terms of the even-parity spherical harmonics,
\bea
{\boldsymbol{\xi}} =  \left( {\boldsymbol{dr}} W_1 + {\boldsymbol{d\th}} V \prt_\th + {\boldsymbol{d\ph}} V \prt_\ph \right) Y_{lm} ,
\label{xi1}
\eea   
where $\{ {\boldsymbol{dr}}, \, {\boldsymbol{d\th}}, \, {\boldsymbol{d\ph}} \}$ is the basis for covariant components of vectors and tensors in the spherical coordinates $\left( r, \th, \ph\right)$, $W_1$ and $V$ are functions depending on $r$ only, and $Y_{lm}$ are the usual spherical harmonics. Note that $\left( {\boldsymbol{d\th}} \prt_\th + {\boldsymbol{d\ph}} \prt_\ph \right) Y_{lm}$ comes from the even-parity vector spherical harmonics \cite{Thorne:1980ru}
\bea
{\boldsymbol{Y}}^{E}_{lm} &:=& \frac{r}{\sqrt{l(l+1)}} \nabla Y_{lm} 
\nonumber \\
&=& \frac{r}{\sqrt{l(l+1)}} \left( {\boldsymbol{d\th}} \prt_\th + {\boldsymbol{d\ph}} \prt_\ph \right) Y_{lm} . 
\label{yelm}
\eea  
Using ${\boldsymbol{\xi}}$, the perturbations of the density, the pressure, and the velocity are 
\bea
&& \de \rh = -\rh \nabla \cdot {\boldsymbol{\xi}} - \left( {\boldsymbol{\xi}} \cdot \nabla \right) \rh ,
\nonumber \\
&& \de p = - \Ga_1 p \nabla \cdot {\boldsymbol{\xi}} - \left( {\boldsymbol{\xi}} \cdot \nabla \right) p ,
\nonumber \\
&& \de {\boldsymbol{v}} = \prt_t {\boldsymbol{\xi}} + \left( {\boldsymbol{v}} \cdot \nabla \right) {\boldsymbol{\xi}} - \left( {\boldsymbol{\xi}} \cdot \nabla \right) {\boldsymbol{v}} , 
\label{fluidpert1}
\eea
where $\Ga_1$ is the adiabatic index for fluid elements under the perturbations. For barotropic EOSs where $p$ depends solely on $\rh$, one has
\bea
\Ga_1 = \frac{\rh}{p}\frac{dp}{d\rh} .
\eea      
Especially, for the polytropic EOSs
\bea
p = \al \rh^{1+\frac{1}{n}},
\label{polytrop1}
\eea 
where $\al$ and $n$ are constants, we have $\Ga_1 = 1+1/n$ with $n$ being the polytropic index.

Using Eq.~\rf{xi1} and Eq.~\rf{fluidpert1} to write out the linear order of Eq.~\rf{basiceqs1}, one finds that the variables $W_1$ and $V$ can be expressed in terms of $\de \Ph$, so that there is only one ordinary differential equation (ODE) for $\de \Ph$ to be solved:
\bea
\de \Ph'' + \frac{2}{r} \de\Ph' + \left( 4\pi \rh \frac{\rh'}{p'} - \frac{l(l+1)}{r^2} \right) \de \Ph = 0 .
\label{perteq1}
\eea
Note that Eq.~\rf{perteq1} is supposed to be valid at the linear order, so $\rh, \, p$ and their radial derivatives take the zeroth order values. Equation~\rf{perteq1} has the simple solution $C_l r^l + D_l r^{-l-1}$ outside the star, where $C_l$ and $D_l$ are constants. Interpreting the $r^l$ terms as an external tidal field and the $r^{-l-1}$ terms as the multipolar response of the star to the external tidal field, the tidal Love numbers are defined as
\bea
k_l := \frac{1}{2} \frac{1}{R^{2l+1}} \frac{D_l}{C_l} ,
\label{tidallove}
\eea
where $R$ is the radius of the star. Focusing on the lowest tidal term which is proportional to $r^2$, it causes a quadrupole term as the response, namely that it is the $l=2$ case. In this case, the tidal deformability is defined as
\bea
\la_2 := \frac{1}{3} \frac{D_2}{C_2} = \frac{2k_2}{3} R^5 .
\label{tidalla}
\eea

The fluid variables and the Newtonian potential need to be finite at the center of the star. This turns out to be a condition that fixes the ratios between $C_l$ and $D_l$, so that $k_l$ can be calculated uniquely given a solution of $\de \Ph$ inside the star. In fact, an expansion analysis of Eq.~\rf{bgeq1} and Eq.~\rf{perteq1} leads to
\bea
&& \Ph^{(0)\, \prime} \approx -\frac{4\pi}{3} r \rh_0 ,
\nonumber \\
&& p^{(0)} \approx p_0 ,
\nonumber \\
&& \de \Ph \approx A_l r^l,
\label{centerapp1}
\eea
at $r \rightarrow 0$, where $\rh_0$ and $p_0$ are the density and the pressure at the center of the star, and $A_l$ are constants that simply scale the solution of $\de \Ph$. For a given EOS, $\rh_0$ and $p_0$ are related. Therefore the solutions to Eq.~\rf{bgeq1} and Eq.~\rf{perteq1} are controlled by one free parameter. 

Given a barotropic EOS, one can numerically integrate Eq.~\rf{bgeq1} and Eq.~\rf{perteq1} from a tiny $r$ where the approximation in Eq.~\rf{centerapp1} is valid to a radius where $p$ vanishes to obtain the solution inside the star. By matching $\de \Ph$ and $\de \Ph'$ at the surface of the star to the exterior solution $C_l r^l + D_l r^{-l-1}$, $C_l$ and $D_l$ can be determined so that $k_l$ can be calculated. As for the moment of inertia, it is obtained simply through the integral
\bea
I = \int \left( r \sin\th \right)^2 \rh^{(0)} \, d^3x .
\eea

\subsection{General relativity}
\label{sec:IIb}

The perturbation theory in GR to calculate the moment of inertia and the tidal deformability of a static spherical star is well-established in the literature; e.g., see Refs.~\cite{Hartle:1967he,Hinderer:2007mb,Damour:2009vw,Binnington:2009bb}. To compare with the formulae in Newtonian gravity parallelly, we briefly review the GR counterpart.  

The basic equations to start with are the Einstein field equations
\bea
G_\mn = 8\pi T_\mn ,
\eea
with the energy-momentum tensor of the fluid 
\bea
T_\mn = \left( \ep + p \right) u_\mu u_\nu + p g_\mn ,
\eea
where $\ep, \, p$ and $u_\mu$ are the proper energy density, the proper pressure, and the four-velocity of the fluid elements. The energy-momentum conservation equation $D^\mu T_\mn = 0$, where $D_\mu$ is the covariant derivative, is a consequence of the Einstein field equations. The variables are set up to be the static spherical background configuration plus perturbations,
namely  
\bea
&& g_\mn = g_\mn^{(0)} + \de g_\mn, 
\nonumber \\
&& \ep = \ep^{(0)} + \de \ep, 
\nonumber \\
&& p = p^{(0)} + \de p,
\nonumber \\
&& u_\mu = u_\mu^{(0)} + \de u_\mu .
\label{varexp2}
\eea 
Using the spherical ansatz 
\bea
g^{(0)}_\mn dx^\mu dx^\nu = g^{(0)}_{tt} dt^2 + \left( 1-\frac{2m}{r} \right)^{-1} dr^2 + r^2 d\Om^2 ,
\eea
for the background metric, one obtains from the Einstein field equations the Tolman-Oppenheimer-Volkoff (TOV) equation
\bea
&& m' = 4\pi r^2 \ep^{(0)} ,
\nonumber \\
&& p^{(0)\, \prime} = -\left(\ep^{(0)}+p^{(0)}\right) \frac{m+4\pi r^3 p^{(0)}}{r(r-2m)} ,
\eea 
where the prime denotes the derivative with respect to $r$.

The perturbations are decomposed into spherical harmonic modes to naturally satisfy the angular dependence in the Einstein field equations. In addition to the even-parity vector spherical harmonics in Eq.~\rf{yelm}, we also need the odd-parity vector spherical harmonics \cite{Thorne:1980ru}     
\bea
{\boldsymbol{Y}}^{B}_{lm} &:=& \frac{1}{\sqrt{l(l+1)}} {\boldsymbol{r}} \times \nabla Y_{lm} 
\nonumber \\
&=& \frac{r}{\sqrt{l(l+1)}} \left( -\frac{{\boldsymbol{d\th}}}{\sin\th} \prt_\ph + {\boldsymbol{d\ph}} \sin\th \prt_\th \right) Y_{lm} . 
\label{yblm}
\eea 
In the Regge-Wheeler gauge \cite{Regge:1957td}, the even-parity perturbations of the metric are 
\bea
\de g_\mn = e^{i\si t}
\begin{pmatrix}
-g^{(0)}_{tt} H_0 & H_1 & 0 & 0  \\
{\rm sym} & g_{rr}^{(0)}H_2 & 0 & 0  \\
0 & 0 & r^2 K & 0 \\
0 & 0 & 0 & r^2 \sin^2\th K
\end{pmatrix} Y_{lm},
\label{evengpertsim}
\eea
%\ATO{I think we are missing a ``sym'' at the (0,1) position in the previous matrix.}
while the odd-parity perturbations of the metric are
\bea
\de g_\mn = e^{i\si t}
\begin{pmatrix}
0 & 0 & -h_0 \frac{1}{\sin\th} \prt_\ph & h_0 \sin\th \prt_\th  \\
{\rm sym } & 0 & -h_1 \frac{1}{\sin\th} \prt_\ph & h_1 \sin\th \prt_\th  \\
{\rm sym } & {\rm sym } & 0 & 0 \\
{\rm sym } & {\rm sym } & {\rm sym } & 0
\end{pmatrix} Y_{lm},
\label{oddgpertsim}
\eea
where $H_0,\, H_1,\, H_2, \, K$ and $h_0,\, h_1$ are functions of $r$, and $\si$ is the oscillation frequency of the mode. The perturbations corresponding to a static tidal field and a slowly rigid rotation are time-independent, so the limit $\si\rightarrow 0$ is taken once the equations for the perturbation variables are obtained. The symbol ``sym'' means the symmetric property $\de g_\mn = \de g_{\nu\mu}$.

For the perturbations in the fluid sector, the fundamental variable is the Lagrangian displacement $\xi_\mu$ for the perturbed fluid elements. It takes the form
\bea
\xi_\mu = e^{i\si t} \left( W_0, W_1, V \prt_\th, V\prt_\ph \right) Y_{lm},  
\eea 
for the even parity, and 
\bea
\xi_\mu = e^{i\si t} \left( 0, 0, -U \frac{1}{\sin\th} \prt_\ph, U \sin\th\prt_\th \right) Y_{lm},
\eea
for the odd parity, where $W_0,\, W_1, \, V$ and $U$ are functions of $r$. The perturbations of the energy density, the pressure, and the four-velocity can be expressed in terms of $\de g_\mn$ and $\xi_\mu$ via \cite{Andersson:2006nr}
\bea
&& \De \ep = -\frac{1}{2} (\ep+p) \left( g^\mn + u^\mu u^\nu \right) \De g_\mn,
\nonumber \\
&& \De p = -\frac{1}{2} \Ga_1 p \left( g^\mn + u^\mu u^\nu \right) \De g_\mn ,
\nonumber \\
&& \De u^\mu = \frac{1}{2} u^\mu u^\al u^\be \De g_{\al\be}.
\label{fluidpert2}
\eea 
Note that those are Lagrangian perturbations. The Eulerian perturbations that are directly used in the Einstein field equations can be obtained via $\De = \de + {\cal L}_{\xi}$, where ${\cal L}_{\xi}$ is the Lie derivative along the vector $\xi_\mu$. For example,
\bea
{\cal L}_{\xi} u^\mu = \xi^\al D_\al u^\mu - u^\al D_\al \xi^\mu .
\eea
In Eq.~\rf{fluidpert2}, we have used
\bea
\Ga_1 = \frac{\ep+p}{p} \frac{dp}{d\ep} ,  
\eea
which is the relativistic version of the adiabatic index for fluid elements under the perturbations. We take
\bea
p = \al \ep^{1+\frac{1}{n}}
\label{polytrop2}
\eea 
as the relativistic version of the polytropic EOSs, so $\Ga_1 = (1+p/\ep)(1+1/n)$ in this case.

Substituting the perturbations of the metric and the perturbations in the fluid sector into the Einstein field equations and the energy-momentum conservation equation, one obtains ODEs for the variables $H_0, \ H_1, \, H_2, \, K, \, h_0, \, h_1$ and $W_1, \, V, \, U$. We focus on the time-independent case to calculate the tidal deformability with the $l=2$ even-parity perturbation and the moment of inertia with the $l=1$ odd-parity perturbation.

With $\si=0$, the even-parity perturbation equations can be simplified to a single second-order ODE for $H_0$,  
\bw
\bea
&& H_0''+ \frac{2(r-m) - 4\pi r^3 (\ep-p)}{r\left( r-2m\right)} H_0' - \frac{ l(l+1) m }{r (r-2 m) \left(m+4 \pi  r^3 p\right)} H_0  - \frac{4 m^3}{r^2 (r-2 m)^2 \left(m+4 \pi  r^3 p\right)} H_0
\nonumber \\
&&  - \frac{4 \pi  r^3 \left(r \ep'+ l(l+1)p +64 \pi ^2 r^4 p^3-20 \pi  r^2 \ep p - 36 \pi  r^2 p^2\right)}{(r-2 m)^2 \left(m+4 \pi  r^3 p\right)} H_0
\nonumber \\
&& -\frac{4\pi r^2 m \left( 40 \pi  r^2 \ep p + 120 \pi r^2 p^2 - 5\ep - \left(2l^2+2l+9\right) p -4 r \ep' \right) + 8\pi r m^2 \left(2r \ep'+5\ep +15p\right)}{(r-2 m)^2 \left(m+4 \pi  r^3 p\right)} H_0  = 0 .
\label{greveneq}
\eea
\ew
The physical solution behaves as
\bea
H_0 \approx A_l r^l,
\eea
at the center of the star, and
\bea
H_0 \approx C_l r^l + D_l r^{-l-1},
\eea
at infinity. By numerically solving Eq.~\rf{greveneq} from the center of the star to a large enough radius, one can extract $C_l$ and $D_l$ from the solution, and hence calculate the tidal Love numbers using the same definition as in Eq.~\rf{tidallove} and the tidal deformability using the same definition as in Eq.~\rf{tidalla}. 

The odd-parity perturbation equations are much simpler than the even ones. In fact, the equation for the fluid variable $U$ is algebraic. For the time-independent case, we have 
\bea
U \rightarrow \frac{i}{\si} \sqrt{\frac{4\pi}{3}} \, \tilde \Om r^2 ,
\eea
where $\tilde \Om$ is a constant, at the limit $\si \rightarrow 0$, so that the angular velocity of the star is 
\bea
\Om = \sqrt{\frac{2l+1}{3}} \, \tilde \Om \frac{d}{d\cos\th}P_l(\cos\th) .
\eea
The rotation of the star is rigid for $l=1$ and differential for $l \ge 2$. The equation for $h_0$ in the time-independent case is
\bea
&& h_0'' - \frac{8\pi r^2 (\ep+p)}{2 (r-2 m)} h_0' - \frac{8\pi r^3 (\ep+p)+ l\left(l+1\right)r-4 m}{r^2 (r-2 m)} h_0
\nonumber \\
&& + \frac{ \sqrt{\frac{\pi }{3}}\, 32 \pi r^3 (\ep+p)}{r-2 m} \tilde \Om = 0 .
\label{groddeq}
\eea 
At the center of the star, the physical solution behaves as 
\bea
h_0 \approx  a_l r^{l+1} + h_{0 \rm in} ,
\label{groddapp}
\eea
where $a_l$ are constants, and $h_{0 \rm in}$ is a particular solution to Eq.~\rf{groddeq} and proportional to the constant $\tilde \Om$. At infinity, the physical solution behaves as
\bea
h_0 \approx d_l \, r^{-l} ,
\label{groddasym}
\eea 
where $d_l$ are constants. Numerical integrations start near $r=0$ where Eq.~\rf{groddapp} is valid. By setting $a_l=1$ and adjusting $\tilde \Om$ to a suitable value, the asymptotic behavior in Eq.~\rf{groddasym} can be achieved. Especially, for $l=1$, the asymptotic behavior in Eq.~\rf{groddasym} means that the metric component $g_{t\ph}$ behaves as
\bea
g_{t\ph} \approx -\frac{2J}{r} \sin^2\th ,
\eea
at infinity, where $J=\sqrt{3} \,d_1/\sqrt{16\pi}$ is the angular momentum of the spacetime. Therefore, the moment of inertia can be calculated via
\bea
I = \frac{J}{\tilde \Om} = \sqrt{\frac{3}{16\pi}} \frac{d_1}{\tilde \Om},
\eea
once $d_1$ and $\tilde \Om$ are known. 

In practice, there is a trick to solve Eq.~\rf{groddeq} for $l=1$ only. A change of variable 
\bea
h_0 = - \sqrt{\frac{4\pi}{3}} \, r^2 (\om-\tilde \Om) ,
\eea
simplifies Eq.~\rf{groddeq} to a homogeneous equation of $\om$ for $l=1$. And outside the star, $\om$ has the simple solution
\bea
\om = c + \frac{d}{r^3},
\eea  
where $c$ and $d$ are constants. For $h_0$ and $g_{t\ph}$ to have the correct asymptotic behavior, one finds
\bea
&& c = \tilde \Om ,
\nonumber \\
&& d = -2 J .
\eea
Therefore, by numerically solving $\om$ inside the star and matching $\om$ and $\om'$ at the surface of the star, one obtains $c$ and $d$, and hence $I=-d/(2c)$.

%%%%%%%%%%%%%%%%%%%%%%%%%%%%%%%%%%%%%%%%%%%%%%%%%%%%%%%%%%%%%%%%%%%%%%%%%%%%%%%%%%%%
\begin{figure}[ht]
%\captionsetup{justification=raggedright,singlelinecheck=false}
  \includegraphics[width=0.9\linewidth]{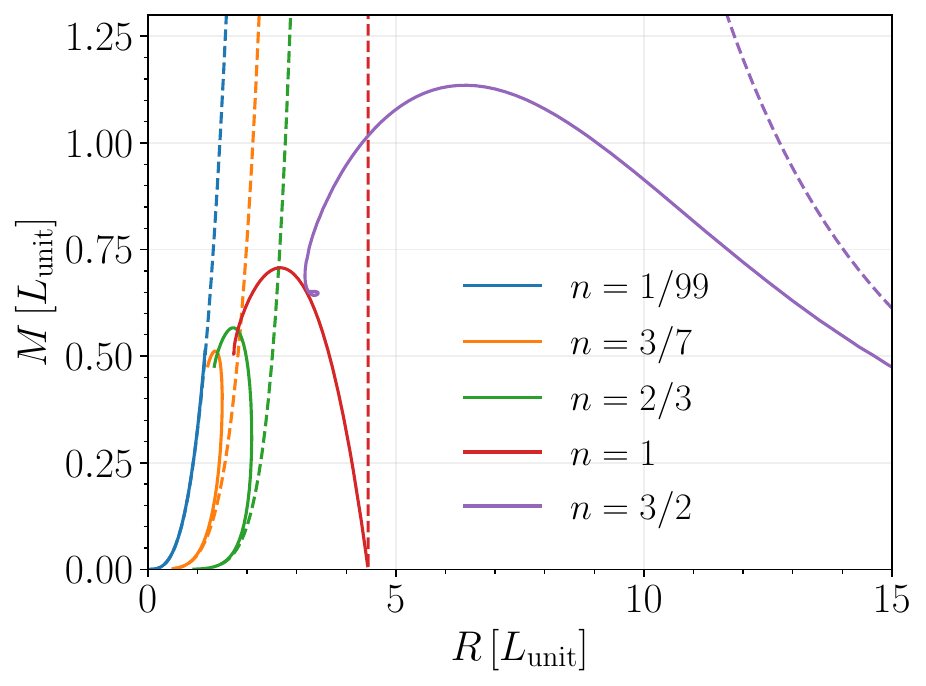}
  \caption{The mass-radius relation for polytropic stars. The dashed lines are the Newtonian results, and the solid lines are the results using GR. Notice that the Newtonian lines extend to infinite mass.}
\label{fig1}
\end{figure}
%%%%%%%%%%%%%%%%%%%%%%%%%%%%%%%%%%%%%%%%%%%%%%%%%%%%%%%%%%%%%%%%%%%%%%%%%%%%%%%%%%%

%%%%%%%%%%%%%%%%%%%%%%%%%%%%%%%%%%%%%%%%%%%%%%%%%%%%%%%%%%%%%%%%%%%%%%%%%%%%%%%%%%%%
\begin{figure}[ht]
%\captionsetup{justification=raggedright,singlelinecheck=false}
  \includegraphics[width=0.9\linewidth]{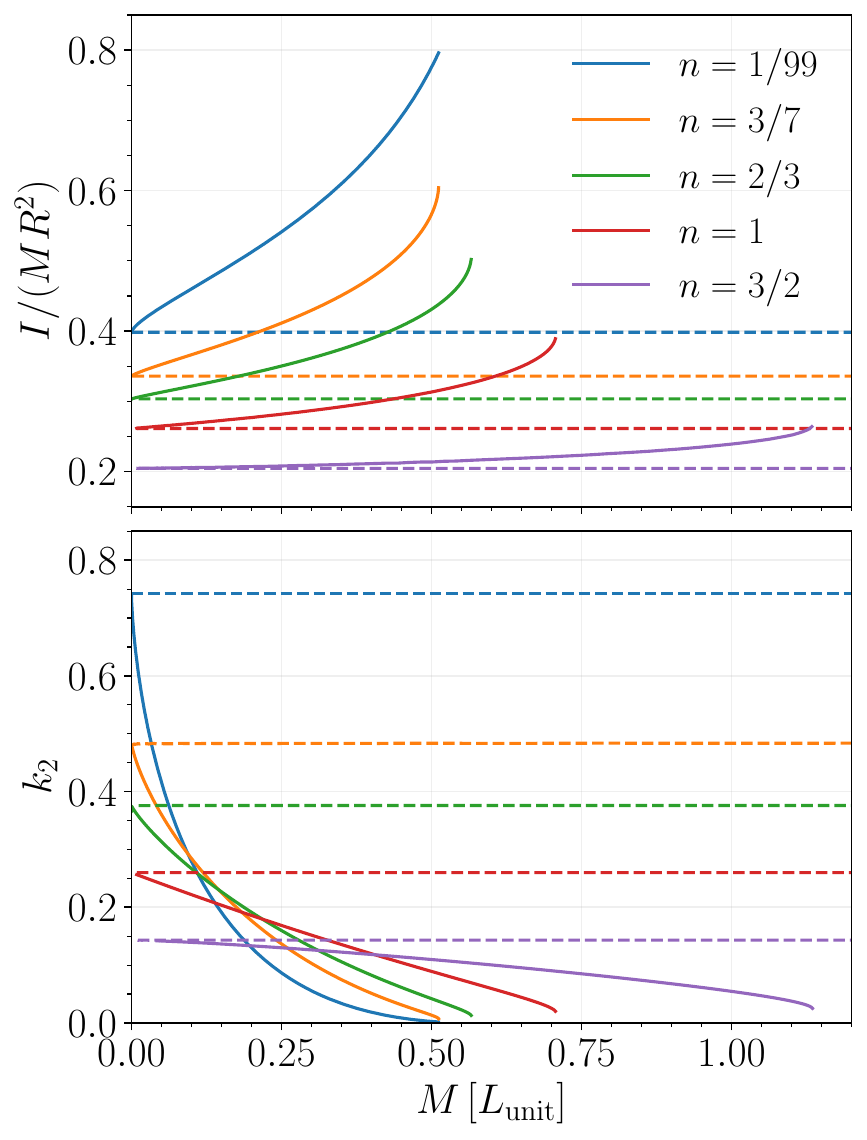}
  \caption{Upper panel: the moment of inertia factor $I/(MR^2)$ versus the mass of the polytropic star. Lower panel: the tidal Love number $k_2$ versus the mass of the polytropic star. The dashed lines are the Newtonian results, and the solid lines are the results using GR. Notice that the Newtonian results do not change with the mass of the star.}
\label{fig2}
\end{figure}
%%%%%%%%%%%%%%%%%%%%%%%%%%%%%%%%%%%%%%%%%%%%%%%%%%%%%%%%%%%%%%%%%%%%%%%%%%%%%%%%%%%

%%%%%%%%%%%%%%%%%%%%%%%%%%%%%%%%%%%%%%%%%%%%%%%%%%%%%%%%%%%%%%%%%%%%%%%%%%%%%%%%%%%%
\begin{figure}[ht]
%\captionsetup{justification=raggedright,singlelinecheck=false}
  \includegraphics[width=0.9\linewidth]{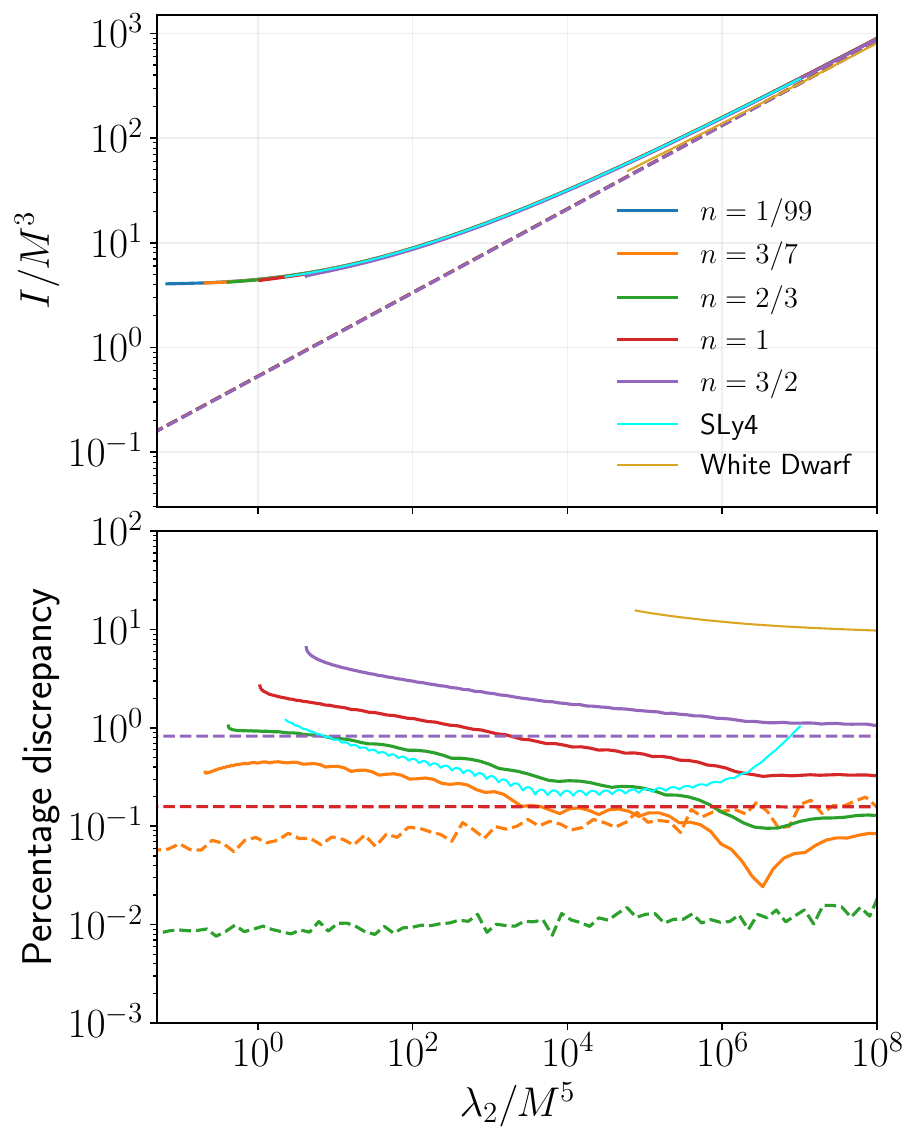}
  \caption{Upper panel: the I-Love universal relation in Newtonian gravity (dashed lines) and in GR (solid lines). Lower panel: Relative discrepancies between each line and the $n=1/99$ lines in Newtonian gravity and in GR, correspondingly. The results using the SLy4 EOS, which is a realistic EOS for NSs, and the zero-temperature EOS for white dwarfs are also plotted for comparison. }
\label{fig4}
\end{figure}
%%%%%%%%%%%%%%%%%%%%%%%%%%%%%%%%%%%%%%%%%%%%%%%%%%%%%%%%%%%%%%%%%%%%%%%%%%%%%%%%%%%

\subsection{Numerical results}
\label{sec:IIc} 

Using the above formulae, we calculate the $l=2$ tidal deformability and the moment of inertia for polytropic stars in both Newtonian gravity and GR. 

First, as a preparation for calculating the tidal deformability and the moment of inertia, masses and radii of the background spherical stars are calculated using polytropic EOSs with different values of indices. Figure~\ref{fig1} presents the results. 
%\ATO{It might be more effective to introduce and position the figures one by one when discussing their contents, as the current approach appears dense and harder to follow.}
The following length parameter has been used as the unit of the mass and radius of the polytropic stars: 
\bea
L_{\rm unit} = \frac{1}{\sqrt{4\pi}} \al^{n/2},
\eea
where $\al$ is the dimensional coefficient, and $n$ is the polytropic index in Eq.~\rf{polytrop1} for Newtonian gravity and Eq.~\rf{polytrop2} for GR. 

One observation from Fig.~\ref{fig1} merits addressing: the Newtonian mass-radius relations extend indefinitely as the central density and pressure approach infinity, meaning there is no maximal mass for Newtonian stars with polytropic EOSs, while the mass-radius relations in GR have maximal masses which mark critical points where the solutions start to be unstable. Note that in the following Figs.~\ref{fig2}-\ref{fig7}, the GR results have been truncated at the maximal-mass points and the unstable solutions are not shown for the clearness of the plots.

Then, in Fig.~\ref{fig2}, results of the moment of inertia and the tidal deformability are converted to the moment of inertia factor $I/(MR^2)$ and the tidal Love number $k_2$, and plotted against the mass of the star. We point out that for a given polytropic index $n$, the moment of inertia factor $I/(MR^2)$ and the tidal Love number $k_2$ remarkably remain constant as the mass of the Newtonian star varies.

Finally, in Fig.~\ref{fig4}, the moment of inertia $I$ and the tidal deformability $\la_2$, both parameterized by the mass of the star, are plotted. It is evident that different polytropic indices result in universal relations between $I/M^3$ and $\la_2/M^5$ in Newtonian gravity and GR separately. While the universal relation in Newtonian gravity meets the universal relation in GR at the higher end of $\la_2/M^5$, which is the regime of the Newtonian limit, they deviate from each other at the lower end of $\la_2/M^5$, corresponding to large masses and compact stars. The deviations between results for different $n$ are smaller in Newtonian gravity than in GR, indicating that the universality of the relation holds more robustly in Newtonian gravity.

To assess the applicability of the polytropic EOS results to real stars, we also calculated $\la_2$ and $I$ for NSs using the tabulated EOS SLy4 \cite{Chabanat:1997qh,Chabanat:1997un} and for white dwarfs using the theoretical EOS under the zero-temperature approximation (see Appendix~\ref{sec:ILwd}). For NSs, the $I/M^3$ versus $\la_2/M^5$ relation derived from EOS SLy4 aligns well with the universal relation for polytropic EOSs. For white dwarfs, there is a discrepancy of about $10\%$ between their $I/M^3$ versus $\la_2/M^5$ relation and the universal relation in GR. Notably, $\la_2$ and $I$ for white dwarfs were calculated within the GR framework, so their $I/M^3$ versus $\la_2/M^5$ relation is compared with the universal relation in GR, even though Fig.~\ref{fig4} shows that the line for white dwarfs is closer to the Newtonian line.

We note the analytical form of the universal relation in Newtonian gravity, obtained from the special case $n \rightarrow 0$. When $n\rightarrow 0$, the polytropic EOS describes stars with uniform density, allowing analytical solutions for Eq.~\rf{bgeq1} and Eq.~\rf{perteq1}. Considering the density discontinuity at the star's surface, one finds $k_2 = 3/4$ \cite{Poisson:2014}. Additionally, $I/(MR^2) = 2/5$ is known for a spherical star with uniform density. Therefore, in this case, we obtain
\bea
\frac{I}{M^3} = \frac{2^{7/5}}{5} \left( \frac{\la_2}{M^5} \right)^{2/5} .
\label{newtheoretic}
\eea

Let us address that the I-Love universal relation is completely nontrivial before we turn to its application. For demonstration, we plot in Fig.~\ref{fig6} $I/R^3$ versus $\lambda_2/R^5$ where $I$ and $\lambda_2$ are parameterized using the radius $R$ rather than the mass $M$, and in Fig.~\ref{fig7} the compactness ${\cal C}:=M/R$ versus $\lambda_2/M^5$. Neither of them is universal. The miraculous I-Love universal relation holds between $I$ and $\la_2$, specifically parameterized by the mass of the star, not between any other seemingly plausible quantities.

%%%%%%%%%%%%%%%%%%%%%%%%%%%%%%%%%%%%%%%%%%%%%%%%%%%%%%%%%%%%%%%%%%%%%%%%%%%%%%%%%%%%
\begin{figure}[ht]
%\captionsetup{justification=raggedright,singlelinecheck=false}
  \includegraphics[width=0.9\linewidth]{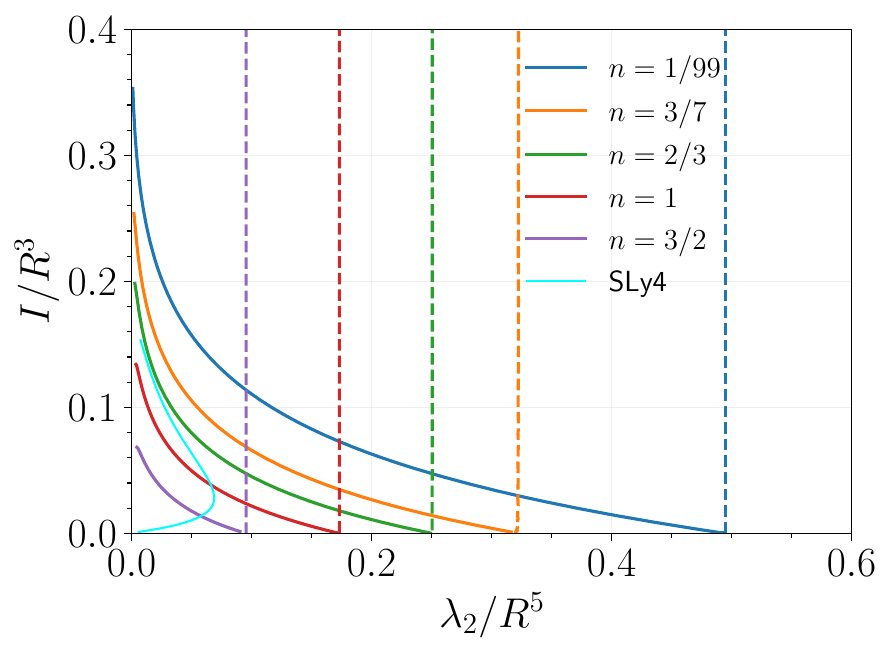}
  \caption{An example of non-universal relation: $I/R^3$ versus $\la_2/R^5$. The dashed lines are the Newtonian results, and the solid lines are the results using GR. The result of NSs (SLy4) is also plotted for comparison. The result of white dwarfs has too small values of $I/R^3$ to show in the plot.}
\label{fig6}
\end{figure}
%%%%%%%%%%%%%%%%%%%%%%%%%%%%%%%%%%%%%%%%%%%%%%%%%%%%%%%%%%%%%%%%%%%%%%%%%%%%%%%%%%%

%%%%%%%%%%%%%%%%%%%%%%%%%%%%%%%%%%%%%%%%%%%%%%%%%%%%%%%%%%%%%%%%%%%%%%%%%%%%%%%%%%%%
\begin{figure}[ht]
%\captionsetup{justification=raggedright,singlelinecheck=false}
  \includegraphics[width=0.9\linewidth]{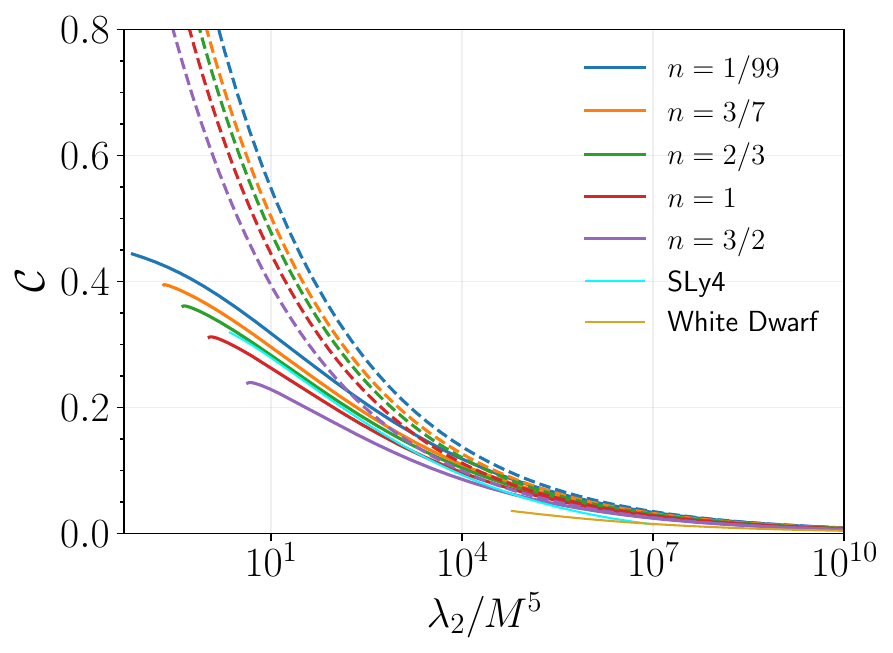}
  \caption{An example of non-universal relation: the compactness ${\cal C}$ versus $\la_2/M^5$. The dashed lines are the Newtonian results, and the solid lines are the results using GR. The results of NSs (SLy4) and white dwarfs are also plotted for comparison. Notice that the compactness of the Newtonian polytropic star extends to infinity.}
\label{fig7}
\end{figure}
%%%%%%%%%%%%%%%%%%%%%%%%%%%%%%%%%%%%%%%%%%%%%%%%%%%%%%%%%%%%%%%%%%%%%%%%%%%%%%%%%%%

%---------------------------------------------------------------------
\subsection{Checking against the planets in the solar system}
\label{sec:checking}
%---------------------------------------------------------------------

%---------------------------------------------------------------------
\begin{table*}
\caption{Measured or theoretically calculated values of the moment of inertia and the tidal Love number for the planets and moons in the solar system.  }
\renewcommand{\arraystretch}{2}
\begin{tabular}{m{2.5cm}m{2.5cm}m{2.5cm}m{2.5cm}m{2.5cm}m{2.5cm}}
\hline\hline
 Object & $M \, [10^{22}\, {\rm kg}]$ & $R\, [10^{6}\, {\rm m}]$ & $I/(MR^2)$ & $k_2$ & Ref. \\
\hline
Mercury & $33.0$ & $2.440$ & $0.333$ & $0.464$  & \cite{2016JGRE..121.1627V,2019GeoRL..46.3625G}  \\
Venus & $4.87\times 10^2$ & $6.052$ & $0.337$ & $0.295$  & \cite{1996GeoRL..23.1857K,2021NatAs...5..676M}   \\
Earth & $5.97\times 10^2$ & $6.378$ & $0.331$ & $0.301$  & \cite{1994AJ....108..711W,2016AcGG...51..493J}   \\
Mars & $64.2$ & $3.396$ & $0.364$ & $0.169$  & \cite{2011Icar..211..401K,2020GeoRL..4790568K}   \\
Jupiter & $1.90\times 10^5$ & $71.49$ & $0.276$ & $0.535$  & \cite{2016ApJ...831...14W,2018AA...613A..32N}   \\
Saturn & $5.68\times 10^4$ & $60.27$ & $0.22$ & $0.379$  & \cite{2018stfc.book...44F,2024AA...684L...3L}   \\
Uranus & $8.68\times 10^3$ & $25.56$ & $0.23$ & $0.319$  & \cite{1995geph.conf....1Y,2021PSJ.....2..222S}   \\
Neptune & $1.02\times 10^4$ & $24.76$ & $0.23$ & $0.29$  & \cite{1995geph.conf....1Y,2024SSRv..220...21J}   \\
\hline
Moon & $7.35$ & $1.737$ & $0.393$ & $0.0242$  & \cite{1996PSS...44.1077W,2014JGRE..119.1546W}   \\
Ceres & $9.47 \times 10^{-2}$ & $0.475$ & $0.36$ & $1.33$  & \cite{2005JGRE..110.5009M,2018Icar..299..430M}   \\
Io & $8.93$ & $1.822$ & $0.377$ & $0.09$  & \cite{2010PSS...58.1381Z,2022Icar..37314737K}   \\
Europa & $4.80$ & $1.561$ & $0.355$ & $0.24$  & \cite{2004jpsm.book..281S,2024Icar..41716120P}   \\
Ganymede & $14.8$ & $2.631$ & $0.311$ & $0.45$  & \cite{2004jpsm.book..281S,2024SSRv..220...54V}   \\
Callisto & $10.8$ & $2.410$ & $0.355$ & $0.33$  & \cite{2003Icar..166..223M,2024SSRv..220...54V}   \\
Titan & $13.5$ & $2.576$ & $0.341$ & $0.62$  & \cite{2012Sci...337..457I,2019Icar..326..123D}   \\
Enceladus & $1.08\times 10^{-2}$ & $0.190$ & $0.335$ & $0.02$  & \cite{2024PSJ.....5...40G}   \\
Rhea & $0.231$ & $0.7644$ & $0.391$ & $1.42$  & \cite{2007GeoRL..34.2202A}   \\ 
\hline
\end{tabular}
\label{planetsdata}
\end{table*}
%---------------------------------------------------------------------

%For NSs, the I-Love universal relation has proved difficult to verify through measurements as there are no simultaneous measurements of $I$ and $\la_2$ for NSs so far. For Newtonian stars, the I-Love universal relation can actually be verified through measurements using the data of the planets and their moons in our solar system. Table~\ref{planetsdata} lists the masses, the radii, the moment of inertia factors, and the tidal Love numbers for the 8 planets in the solar system as well as for some of their moons. Some of the moment of inertia factors and the tidal Love numbers are not exactly direct measurements, but theoretically inferred values from models built upon the current best knowledge of these celestial objects. We use these models for comparison as they are expected to be much more reliable than the simple model of spherical polytropic stars.    

For NSs, verifying the I-Love universal relation through measurements remains
challenging due to the lack of simultaneous observations of $I$ and $\la_2$.
However, for Newtonian stars, the I-Love universal relation can be validated
using measurements from planets and moons within our solar system.
Table~\ref{planetsdata} provides the masses, radii, moment of inertia factors,
and tidal Love numbers for the 8 planets and some of their moons. While some
moment of inertia factors and tidal Love numbers are not directly measured,
they are theoretically inferred from models based on the current best
understanding of these celestial objects. These models are used for comparison,
as they are considered significantly more reliable than the simplistic model of
spherical polytropic stars.

%In Fig.~\ref{fig5}, the data of the moment of inertia and the tidal Love number for the celestial objects listed in Table~\ref{planetsdata} are plotted around the theoretical line of Eq.~\rf{newtheoretic}. The trend of the data aligns well with the theoretical line\ATO{', in particular for planets'}. A linear regression of the data under the logarithmic scale gives

\begin{figure}[ht]
%\captionsetup{justification=raggedright,singlelinecheck=false}
  \includegraphics[width=0.9\linewidth]{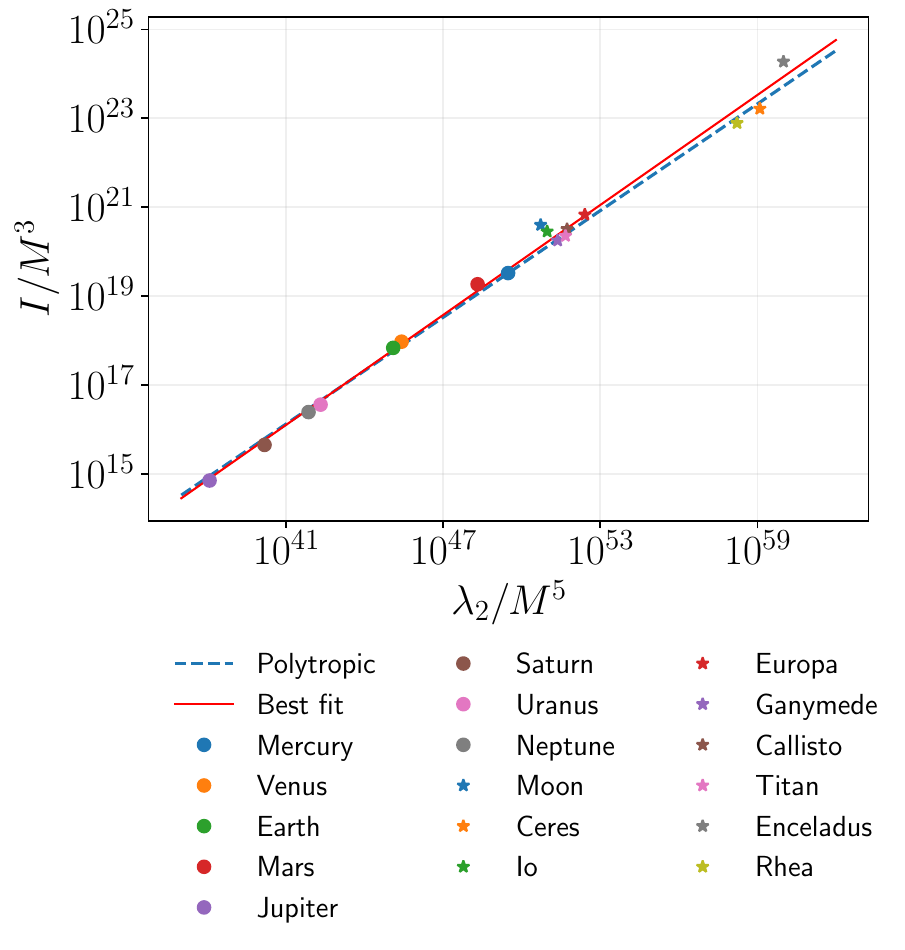}
  \caption{The theoretical I-Love universal relation checked against the measurements of the solar-system celestial bodies.  }
\label{fig5}
\end{figure}

In Fig.~\ref{fig5}, we depict the data for the moment of inertia and tidal Love
number of the celestial objects listed in Table~\ref{planetsdata}, alongside
the theoretical line of Eq.~\rf{newtheoretic}. The trend of the data aligns
well with the theoretical line, particularly for planets. A linear
regression of the data in the logarithmic scale yields:

\bea
\log_{10} \left( \frac{I}{M^3} \right) = 0.412 \log_{10} \left( \frac{\lambda_2}{M^5} \right) - 0.803 ,
\eea
with a R-Squared value of $0.984$. The slope $0.412$ is slightly greater than the theoretical value $2/5$. This is likely because the planets and the moons are slightly oblate due to their rotations or tides from nearby bodies, so they have larger moments of inertia than spherical bodies of the same masses.

%%%%%%%%%%%%%%%%%%%%%%%%%%%%%%%%%%%%%%%%%%%%%%%%%%%%%%%%%%%%%%%%%%%%%%%%%%%%%%%%%%%%
%%%%%%%%%%%%%%%%%%%%%%%%%%%%%%%%%%%%%%%%%%%%%%%%%%%%%%%%%%%%%%%%%%%%%%%%%%%%%%%%%%%

%---------------------------------------------------------------------
\section{Summary}
\label{sec:sum}
%---------------------------------------------------------------------
We have numerically calculated the moment of inertia (I) and the tidal Love number (Love) for polytropic stars in both Newtonian gravity and GR. The I-Love relation varies within $1\,\%$ for different polytropic indices in Newtonian gravity and within $10\,\%$ for different polytropic indices in GR. The results explicitly demonstrate that the I-Love universal relation not only exists for compact stars but also exists for Newtonian stars.   

We have also computed the moment of inertia and the tidal Love number for realistic NSs and white dwarfs to compare with the polytropic stars. We find that the I-Love relation of the realistic NSs falls within $1\,\%$ of the I-Love relation of the $n\rightarrow 0$ polytropic case. For white dwarfs, their I-Love relation has a $10\,\%$ discrepancy away from the $n\rightarrow 0$ polytropic case when comparing to the GR solution; surprisingly, the I-love relation for white dwarfs agrees better with the Newtonian result. 

Finally, we have collected measured and theoretically inferred data of the moment of inertia and the tidal Love number for 17 celestial bodies in our solar system to check the I-Love universal relation obtained for polytropic stars. The trend of the data follows the theoretical line well. The slope of the best-fit line of the data deviates from the theoretical value $2/5$ by only $3\,\%$, showing the adequacy of the theoretical I-Love universal relation in describing real planets and moons. Therefore, the I-Love universal relation provides a valuable tool for exploring exoplanets. Especially when combined with the Darwin–Radau equation \cite{Bourda:2004jxj}, one can compute the moment of inertia and the tidal Love number of the exoplanet if its mass, spin frequency, equatorial radius, and polar radius are known. The moment of inertia and the tidal Love number are closely related to the interior structure and compositions of the planet. Knowing them helps build detailed models of the exoplanet and thus understand it better.

%---------------------------------------------------------------------
\acknowledgments 
%---------------------------------------------------------------------

This work was supported by the China Postdoctoral Science Foundation (2023M741999), and the high-performance computing cluster in the Astronomy Department of Tsinghua University.

%---------------------------------------------------------------------
\appendix
\section{Zero-temperature white dwarfs}
\label{sec:ILwd}
%---------------------------------------------------------------------
Here we append numerical results on the tidal deformability and the moment of inertia of white dwarfs, using the EOS of the zero-temperature degenerated electron gas. The derivation of the EOS can be found, for example, in Refs.~\cite{Shapiro:1983du,chandra1,Poisson:2014}; we write down the parameterized EOS directly for our use. The number density, the energy density, and the pressure of the zero-temperature degenerated electron gas are 
\bea
&& n_e = \frac{m_e^3}{ 3\pi^2 \hbar^3} x^3 ,
\nonumber \\
&& \ep_e = \frac{m_e^4}{8\pi^2 \hbar^3} \left[ x\sqrt{1+x^2} \left( 1+2x^2 \right) + \ln{\left(\sqrt{1+x^2}-x\right)} \right],
\nonumber \\
&& p_e = \frac{m_e^4}{8\pi^2 \hbar^3} \left[ x\sqrt{1+x^2} \left( \frac{2}{3}x^2 - 1 \right) - \ln{\left(\sqrt{1+x^2}-x\right)} \right] ,
\label{degeneratedelectron}
\eea    
where $x:= \sqrt{(E_F/m_e)^2 - 1}$ with $E_F$ the Fermi energy of the electron gas and $m_e$ the electron mass. While the pressure in white dwarfs can be approximated as 
\bea
p\approx p_e,
\label{wdp}
\eea
the energy density in white dwarfs must take the rest energy of the nuclei into account, namely
\bea
\ep \approx \rh_n + \ep_e = \mu_e m_H n_e + \ep_e ,
\label{wdep}
\eea
where $\mu_e \approx 2$ is the mean molecular weight and $m_H=1.660539\times 10^{-27}\, {\rm kg}$ is the atomic mass unit.

\begin{figure}[ht]
%\captionsetup{justification=raggedright,singlelinecheck=false}
  \includegraphics[width=0.9\linewidth]{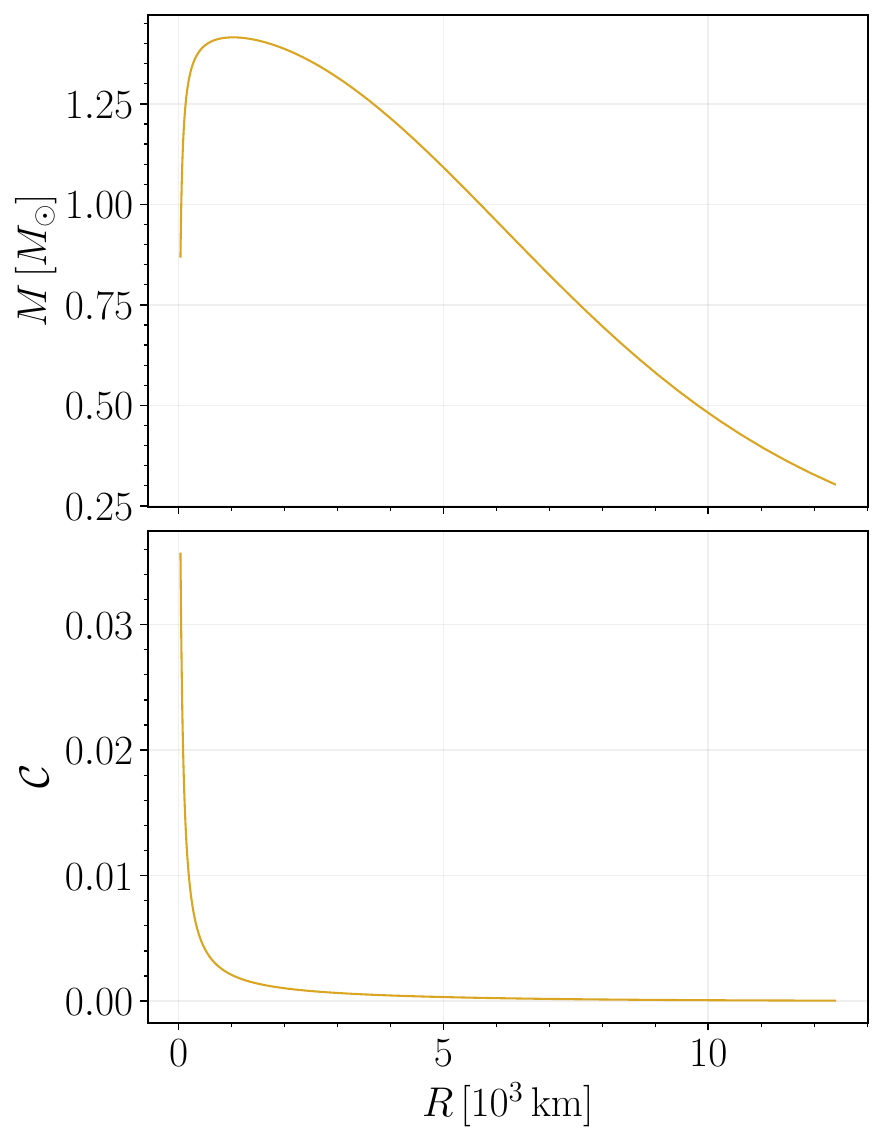}
  \caption{Upper panel: the mass-radius relation of the zero-temperature white dwarfs. Lower panel: the compactness versus the radius of the zero-temperature white dwarfs. }
\label{figwd1}
\end{figure}

\begin{figure}[ht]
%\captionsetup{justification=raggedright,singlelinecheck=false}
  \includegraphics[width=0.9\linewidth]{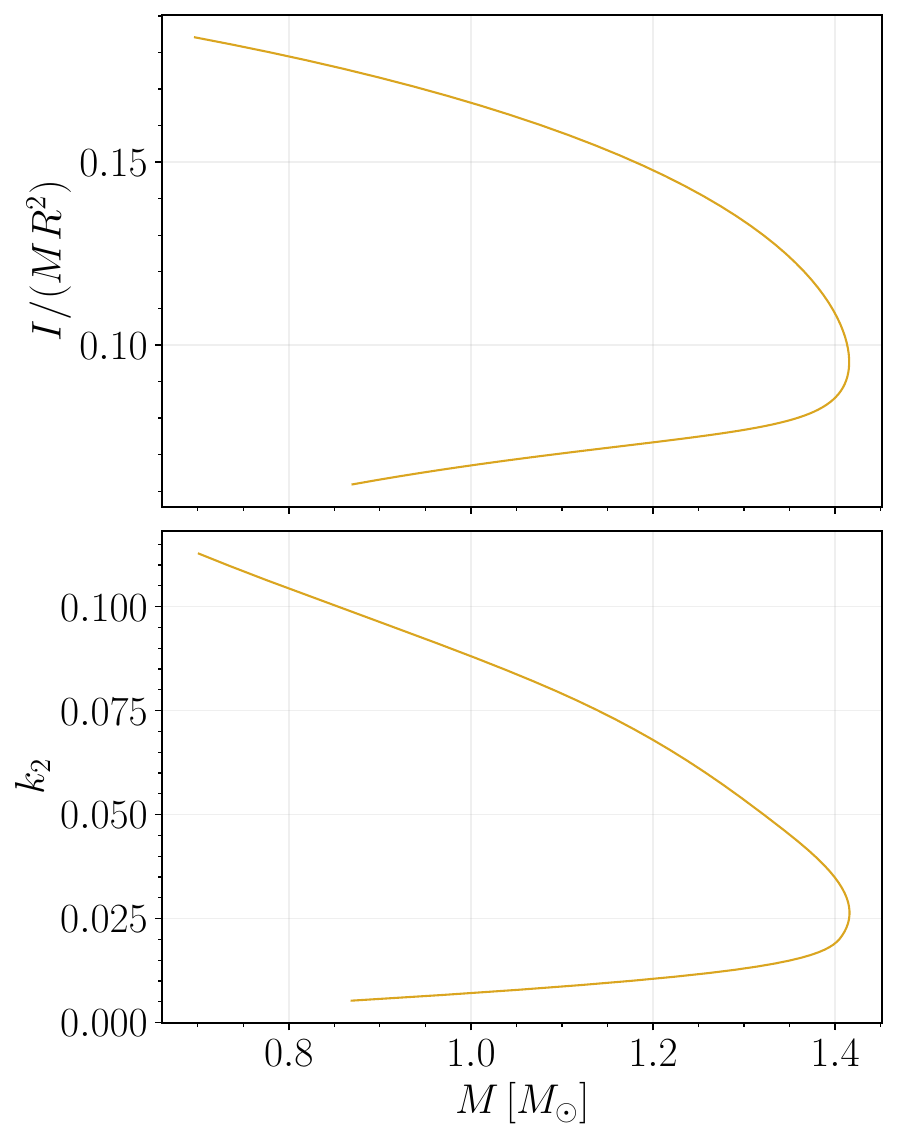}
  \caption{Upper panel: the moment of inertia factor versus the mass of the zero-temperature white dwarfs. Lower panel: the tidal Love number versus the mass of the zero-temperature white dwarfs. }
\label{figwd2}
\end{figure}

\begin{figure}[ht]
%\captionsetup{justification=raggedright,singlelinecheck=false}
  \includegraphics[width=0.9\linewidth]{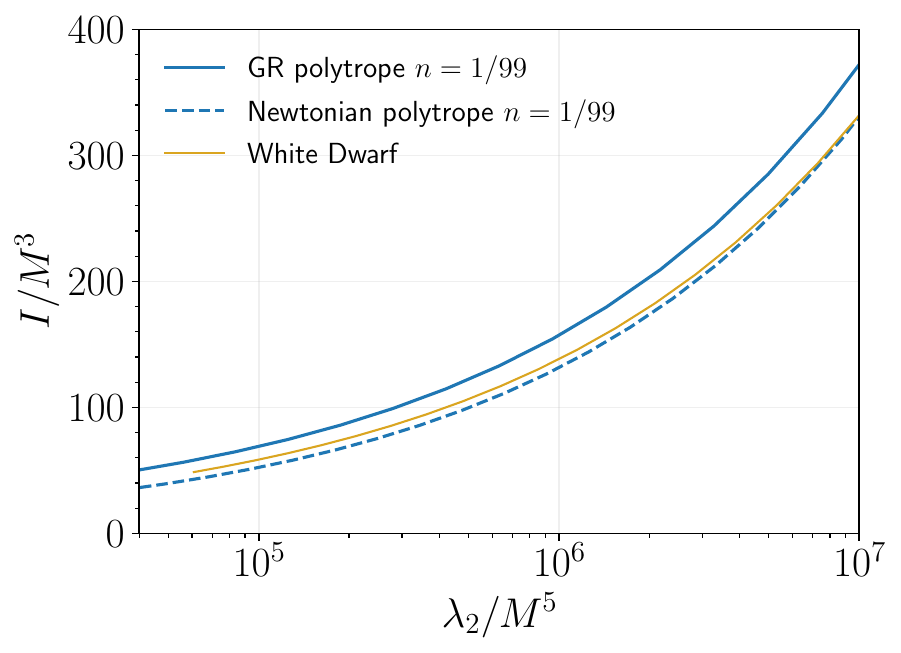}
  \caption{The I-Love relation for the white dwarfs (red solid line) compared with the I-Love universal relation in Newtonian gravity (blue dashed line) and in GR (blue solid line).  }
\label{figwd3}
\end{figure}

Using the same formulae as in Sec~\ref{sec:IIb} but with the EOS given by Eqs.~\rf{wdp} and \rf{wdep}, we have calculated the mass-radius relation, the tidal deformability, and the moment of inertia for white dwarfs. Figures~\ref{figwd1} and \ref{figwd2} show the results. Then in Fig.~\ref{figwd3}, the same I-Love relation plot as in Fig~\ref{fig4} is shown, with the focus now on the white dwarfs and the vertical axis in linear scale.

%%%%%%%%%%%%%%%%%%%%%%%%%%%%%%%%%%%%%%%%%%%%%%%%%%%%%%%%%%%%%%%%%%%%%%%%%%%%%%%%%%%%
%%%%%%%%%%%%%%%%%%%%%%%%%%%%%%%%%%%%%%%%%%%%%%%%%%%%%%%%%%%%%%%%%%%%%%%%%%%%%%%%%%%

%%%%%%%%%%%%%%%%%%%%%%%%%%%%%%%%%%%%%%%%%%%%%%%%%%%%%%%%%%%%%%%%%%%%%%%%%%%%%%%%%%%%
%%%%%%%%%%%%%%%%%%%%%%%%%%%%%%%%%%%%%%%%%%%%%%%%%%%%%%%%%%%%%%%%%%%%%%%%%%%%%%%%%%%

%%%%%%%%%%%%%%%%%%%%%%%%%%%%%%%%%%%%%%%%%%%%%%%%%%%%%%%%%%%%%%%%%%%%%%%%%%%%%%%%%%%%
%%%%%%%%%%%%%%%%%%%%%%%%%%%%%%%%%%%%%%%%%%%%%%%%%%%%%%%%%%%%%%%%%%%%%%%%%%%%%%%%%%%

%---------------------------------------------------------------------
\bibliography{refs}
%---------------------------------------------------------------------

\end{document}